\documentclass[aps,11pt,letterpaper,nofootinbib,amsmath,amssymb]{revtex4}%
\usepackage{latexsym}%
\usepackage{epsf}%
\usepackage{graphicx}%
\usepackage{epsfig}%
\usepackage{graphicx}%
\usepackage{epstopdf}%
\usepackage[dvips,dvipdfmx,hypertex]{hyperref}%

\def\cD{{\mathcal{D}}}

\def\oh{{\overline{h}}}

\def\vr{{\vec{r}}}

\def\oV{{\overline{V}}}
\def\tV{{\widetilde{V}}}

\def\sgn{{\text{sgn}}}
\def\nc{{n_\text{c}}}
\def\rhoeq{{\rho_\text{eq}}}

\def\rhoc{{\rho_\text{c}}}
\def\cH{{\mathcal{H}}}
\def\erf{{\text{erf}}}
\def\cE{{\mathcal E}}
\DeclareMathAlphabet{\mathpzc}{OT1}{pzc}{m}{it}

\newcommand{\beq}{\begin{equation}}
\newcommand{\beqn}{\begin{equation}\nonumber}
\newcommand{\eeq}{\end{equation}}
\newcommand{\bea}{\begin{eqnarray}}
\newcommand{\bean}{\begin{eqnarray}\nonumber}
\newcommand{\eea}{\end{eqnarray}}
\newcommand{\ba}{\begin{align}}
\newcommand{\ea}{\end{align}}

\begin{document}

\title{Gravitationally Bound Bose Condensates with Rotation}
\author{Souvik Sarkar\footnote{\tt sarkarsi@mail.uc.edu}}
\author{Cenalo Vaz\footnote{\tt Cenalo.Vaz@uc.edu}}
\author{L.C.R. Wijewardhana\footnote{\tt Rohana.Wijewardhana@uc.edu}}
\affiliation{Department of Physics, University of Cincinnati, Cincinnati, OH 45221-0011.}

\begin{abstract}

We develop a self-consistent, Gravitoelectromagnetic (GEM) formulation of a slowly rotating, self-gravitating and dilute 
Bose-Einstein condensate (BEC), intended for astrophysical applications in the context of dark matter 
halos. GEM self-consistently incorporates the effects of frame dragging to lowest order in $v/c$ via the Gravitomagnetic 
field. BEC dark matter has attracted attention as an alternative to Cold dark matter (CDM) and Warm dark matter (WDM) for 
some time now. The BEC is described by the Gross-Pitaevskii-Poisson (GPP) equation with an arbitrary potential allowing 
for either attractive or repulsive interactions. Owing to the difficulty in obtaining exact solutions to the GEM equations 
of motion without drastic approximations, we employ the variational method to examine the conditions under which rotating 
condensates, stable against gravitational collapse, may form in models with attractive and repulsive quartic interactions. 
We also describe the approximate dynamics of an imploding and rotating condensate by employing a collective coordinate 
description in terms of the condensate radius.
%\pacs{}
\end{abstract}
\maketitle

\section{Introduction}

The standard ($\Lambda$CDM) model for the matter-energy distribution of the universe posits that the universe 
consists primarily of a cosmological constant ($\Lambda$) plus cold dark matter (CDM). It has been fairly successful on 
scales larger than the typical galactic scale ($\gtrsim$ 50 kpc) but appears to make predictions that are somewhat 
inconsistent with 
observations on smaller scales ($\lesssim$ 50 kpc). In this model, only about 4\% of the matter-energy budget of 
the universe is ordinary baryonic matter as we know it, 22\% is non-baryonic dark matter (DM) and the rest is a mysterious 
form of energy (dark energy, DE), which appears to be reasonably well described by a positive cosmological constant 
\cite{rea98,pea99,eea05}. The model has provided a useful framework for understanding structure formation via 
density fluctuations and has had some success in explaining the power spectrum of the mass fluctuations in the Cosmic 
Microwave Background (CMB) radiation \cite{kea11,hea12,pla15}, the large scale structure \cite{san12,and12} and 
the large scale structure of DM halos \cite{wal09,vik09,new12}. On smaller scales, however, numerical simulations 
\cite{dc91,fp94,mo94,nfw96} predict unobserved cusps in the central dark matter density profiles of galactic halos. On 
the contrary, observations of rotation curves prefer cored distribution profiles {\it i.e.,} having a nearly constant DM 
density near the center \cite{gts07,ohea07,dea09,ns13}. Furthermore, because of the hierarchical growth of structure in the 
CMB models, they predict an excess of low mass sub-halos within the galaxy as well as an excess of massive sub-halos, 
capable of being bright enough to be observed as satellite galaxies \cite{kkvp99,mgea99,swea08,bs09,bkbk11,fw12,wbgkn15}. 
Other problems also indicate that the $\Lambda$CDM model may be in need of tuning: for example, pure disk (bulge free) 
galaxies cannot be simulated in this model \cite{gbea10,kdbc10} and some anomalies between the CMB mass power spectrum 
obtained by the Sloan Digital Sky Survey (SDSS) and that predicted by the CDM paradigm have been found \cite{bbk17}.

A proposed alternative to CDM is a light boson whose mass is small enough that its critical temperature is well above that of 
the CMB. This ensures that a significant fraction of the bosons settle in the ground state and form a BEC  \cite{kaup68,rb69,
wt83,bgz84,imr90,bh07}. If the particle mass is small enough so that its de Broglie wavelength is on the order of the typical 
galactic scale (say $\sim 50$ kpc), smaller scale structure will be suppressed while on the large 
scale it would be virtually indistinguishable from CDM \cite{mst83,prs90,sjs94,hbg00,jg00,pjp00,ab05,phc11,scb14,ds14,lds14,hoea17}. 
Such low masses are not inconceivable, considering that ultra-light bosons (of mass even as low as $\sim 10^{-33}$ ev) are 
predicted by multidimensional cosmology and string theory \cite{hw96,gz97,add99,ar10}. Larger particle masses lead to smaller, 
asteroid sized, stable, BEC structures (one example is the QCD axion), which, in sufficient numbers, could form one component 
of DM \cite{csw86,kmz85,bsk92,sk94,ssk96,kss99,sg01,krs04,k08,bb11,ghp15,esvw14,elsw16,lpt17,phc17,hmea17,sh17}. Of interest 
in either case, but particularly for large halos, is the description of rotating condensates \cite{sds95,mmea99,rds12,ds16}.

In this work, we provide a Gravitoelectromagnetic (GEM) description of gravitationally bound, non-relativistic, rotating BECs. 
It is not our intention to obtain exact solutions of the equations of motion. Instead, we analyze the effects of the rotation 
by applying the variational method. The variational method is widely used in the study of BECs, in condensed matter physics 
\cite{bist97,ps08} and was adopted early in the Boson star literature (eg. \cite{phc11}) as well, owing to the difficulty in 
obtaining solutions of the coupled system of equations. 

In section II, beginning with the relativistic Klein-Gordon field in a weak, axisymmetric gravitational field, we set up the 
effective Gross-Pitaevskii-Poisson (GPP) action, describing a BEC cloud with rotation and including the gravitational action 
appropriate to an axisymmetric spacetime in the weak field approximation. In section III, we describe stable configurations 
by applying the variational method with a single Gaussian Vortex ansatz appropriate to rotating BEC clouds. In section IV we 
examine the problem of vortex collapse and we summarize our results in section V.

\section{Action}

As the candidate field for BEC halos is generally taken to be a light, but not massless, scalar field, we begin with the 
action for a complex scalar field in a curved background spacetime, $S=S_\phi+S_G$, where
\beq
S_\phi = - \int d^4x \sqrt{-g}\left[g^{\alpha\beta}(\nabla_\alpha \phi)^*(\nabla_\beta\phi) + V(|\phi|)\right],
\label{complexscalar}
\eeq
and $V(|\phi|)$ is an arbitrary potential, which we take to be of the form
\beq
V(|\phi|) = \frac{m^2 c^2}{\hbar^2}|\phi|^2 + \frac{2m}{\hbar^2}\tV_1(|\phi|).
\eeq
This action is supplemented by the gravitational action
\beq
S_G = \frac{c^3}{16\pi G} \int d^4 x \sqrt{-g}~ R,
\eeq
linearized about flat space, taking $g_{\mu\nu} = \eta_{\mu\nu} + h_{\mu\nu}$.

\subsection{The GPP Action}

In the weak field approximation \cite{tp10,sc16}, $g_{\mu\nu} = \eta_{\mu\nu}+ h_{\mu\nu}$ where $|h_{\mu\nu}| \ll |\eta_{\mu\nu}|$,
if we set $\sqrt{-g} \approx 1 + \frac 12 h$, where $h=\eta^{\mu\nu}h_{\mu\nu}$ is the trace of $h_{\mu\nu}$, then the scalar 
field action can be written as the sum of three terms,
\beq
S_\phi = S_0[\phi] - \int d^4x\left[\frac 12 h \left(\eta^{\alpha\beta}(\nabla_\alpha \phi)^*(\nabla_\beta\phi)+ V(|\phi|)
\right) - h^{\alpha\beta} (\nabla_\alpha \phi)^*(\nabla_\beta\phi)\right]
\eeq
to linear order in $h_{\mu\nu}$. The zeroeth order action, $S_0[\phi]$, is the action in \eqref{complexscalar}, but on 
a flat background. We will call the trace term $S_1[h,\phi]$ and the second correction term $S_2[h_{\alpha\beta},\phi]$. 
In the non-relativistic limit, the field can be described by a complex wavefunction $\psi$ according to
\beq
\phi = \frac\hbar{\sqrt{2m}} e^{-imc^2t/\hbar}\psi,
\eeq
which, at low temperatures, describes a condensed, $N$-particle state and will be normalized as $\int d^3 \vr |\psi|^2 = N$ in 
what follows. We will also take the scalar field potential to be of the form
\beq
V(|\phi|) = \frac{mc^2}{2}|\psi|^2 + \tV(|\psi|).
\eeq
Expanding the zeroeth order action in the non-relaivistic limit, one finds
\beq
S_0[\phi] \approx \int dt\int d^3\vr \left[\frac{i\hbar}2 \psi^*\overleftrightarrow{\nabla_t}\psi - \frac{\hbar^2}{2m} 
|\nabla\psi|^2 - \tV(|\psi|)\right]
\eeq
where we have dropped terms quadratic in $\dot\psi/c$, using the fact that, in the non-relativistic approximation, the 
difference between the total energy and the rest mass energy is supposed to be small. In the same approximation, 
\beq
S_1[\phi] \approx \int dt \int d^3 \vr~ h \left[\frac{i\hbar}2 \psi^*\overleftrightarrow{\nabla_t}\psi - \frac{\hbar^2}{2m} 
|\nabla\psi|^2 - \tV(|\psi|)\right].
\eeq
Assuming the gravitational field is weak ($|h|\ll 1$), this term may therefore be dropped, as also the last term in the expansion
of $S_2[h_{\alpha\beta},\phi]$,
\bea
S_2[h_{\alpha\beta},\phi] &=& \int d^4x~ h^{\alpha\beta}(\nabla_\alpha \phi)^*(\nabla_\beta\phi) \cr\cr
&=&\int d^4x \left[h^{tt} |\dot\phi|^2 + h^{ti} \left(\dot\phi^*\nabla_i \phi + \nabla_i\phi^* \dot\phi\right) + 
h^{ij}\nabla_{(i}\phi^*\nabla_{j)}\phi\right]\cr\cr
&\approx&\int dt \int d^3\vr \left[\frac 12 mc^4 h^{tt} |\psi|^2 + \frac{i\hbar c^2}2 h^{ti}\left(\psi^*\nabla_i\psi
-\nabla_i \psi^* \psi\right)\right].
\eea
If we call $h^{ti} = -A^i/c^2$, $h^{tt} = -2\Phi_G/c^4$, then this correction term looks like
\beq
S_2[h_{\alpha\beta},\phi] \approx \int dt \int d^3\vr \left[ -m\Phi_G|\psi|^2 - J\cdot A\right],
\eeq
where $\Phi_G$ is the gravitational potential energy (to be obtained from the Einstein equations) and
\beq
J_i = \frac{i\hbar}2 \psi^*\overleftrightarrow{\nabla_i}\psi
\eeq
is the scalar current. Putting everything together, the non-realtivistic, weak field GPP action for the scalar field is
\beq
S_\psi \approx \int d^4x \left[\frac{i\hbar}2 \psi^* \overleftrightarrow{\nabla_t}\psi - \frac{\hbar^2}{2m} |\nabla\psi|^2 
- \tV(|\psi|) - m\Phi_G|\psi|^2 - J\cdot A\right].
\eeq
The third term in the action above represents the self-interaction of the non-relativistic field, the fourth term represents 
its interaction with the gravitational field and the last term is the gravitomagnetic term, which incorporates frame 
dragging.

The non-relativistic action may be put in a more suggestive form if we define $A_\mu = (-\Phi_G,A_i)$ and the 
``covariant'' derivative
\beq
D_\mu \psi = \left(\nabla_\mu - \frac{im}{\hbar} A_\mu\right)\psi.
\eeq
Then, 
\beq
\frac{i\hbar}2\psi^* \overleftrightarrow{D_t} \psi = \frac{i\hbar}2\psi^* \overleftrightarrow{\nabla_t}\psi + 
m A_t |\psi|^2 = \frac{i\hbar}2 \psi^* \overleftrightarrow{\nabla_t}\psi - m \Phi_G |\psi|^2
\eeq
and 
\beq
-\frac{\hbar^2}{2m}(D_i\psi)^* (D^i\psi) = -\frac{\hbar^2}{2m} |\nabla\psi|^2 - J \cdot A 
\eeq
upon dropping terms that are quadratic in the gravitational field. Thus our action reads,
\beq
S_\psi \approx \int d^4x \left[\frac{i\hbar}2 \psi^* \overleftrightarrow{D_t}\psi - \frac{\hbar^2}{2m} |D_i\psi|^2 
- \tV(|\psi|)\right],
\eeq
so, in the weak field limit, the interaction of the scalar field with the gravitational field has the (well-known) form 
of an ``electromagnetic'' coupling.

\subsection{The Gravitational Action}

Turning to the gravitational part of the action, we first examine the Einstein equations of motion to determine the 
general form of the metric. It is convenient to work in the harmonic gauge, defining $\oh_{\mu\nu} = h_{\mu\nu} - 
\frac 12 \eta_{\mu\nu}h$ and imposing the condition ${\oh_{\mu\nu}}^{,\nu} = 0$. The Einstein tensor is 
$G_{\mu\nu} = \frac 12 \Box \oh_{\mu\nu}$ and Einstein's equations are
\beq
\Box \oh_{\mu\nu} = \frac{16\pi G}{c^4} T_{\mu\nu}
\eeq
The gauge condition, ${\oh_{\mu\nu}}^{,\nu}=0$ tells us that the stress tensor is conserved on a flat background, so 
$T_{\mu\nu}$ is to be evaluated for the field on a flat background,
\beq
T_{\mu\nu} = \nabla_\mu\phi^* \nabla_\nu\phi + \nabla_\nu\phi^* \nabla_\mu\phi + \eta_{\mu\nu} \mathcal{L}.
\eeq
In the non-relativistic limit, keeping only leading order terms, we find
\bea
T_{tt} &\approx& mc^4|\psi|^2 \cr
T_{ti} &\approx& c^2J_i\cr
T_{ij} &\approx& 0
\eea
This shows that we can take $\oh_{ij} = 0$. In this case, $\oh=-h = -\oh_{tt}/c^2$. If we take $\oh_{tt} = -4\Phi_G$
then $\oh = 4\Phi_G/c^2$ and $h_{tt} = -2\Phi_G$. The remaining metric coefficients are
\beq
h_{ij} = \oh_{ij} - \frac 12 \eta_{ij} \oh = -\frac{2\Phi_G}{c^2}\delta_{ij},~~ h_{ti} = \oh_{ti} = -A_i
\eeq
and the line element,
\beq
ds^2 = c^2(1+2\Phi_G/c^2)dt^2 +2A_i dt dx^i - (1-2\Phi_G/c^2)\delta_{ij} dx^i dx^j,
\label{linelement}
\eeq
is subject to the gauge conditions,
\beq
{\oh_{t\mu}}^{,\mu} = \frac{4\dot\Phi_G}{c^2} - \nabla\cdot A = 0,~~ {\oh_{i\mu}}^{,\mu} = \frac 1{c^2}
\dot A_i = 0.
\eeq
In the non-relativistic limit, we may ignore the term $\dot\Phi_G/c^2$, so our gauge conditions become
$\dot A_i = 0 = \nabla\cdot A$.

To set up the total action it is convenient to compare the line element in \eqref{linelement} to the standard line 
element of Arnowitt, Deser and Misner (ADM) \cite{adm62}
\beq
ds^2 = N^2 dt^2 - N_i dt dx^i - \gamma_{ij} dx^i dx^j
\eeq
and write the bulk Lagrangian for the gravitational field in terms of the lapse function, $N$, the shift, 
$N_i$, the first fundamental form, $\gamma_{ij}$, the intrinsic curvature scalar of spatial hypersurfaces, 
${}^{(3)}R $, and the extrinsic curvature (of spatial hypersurfaces), $K_{ij}$,
\beq
\mathfrak{L}_G = \frac{c^3N}{16\pi G} \sqrt{\gamma} \left[{}^{(3)}R + K_{ij}K^{ij} - K^2\right]
\label{gravlagden}
\eeq
where $K$ is the trace of $K_{ij}$. This gives us the lapse, shift and first fundamental form respectively 
in linear approximation,
\bea
&&N \approx c\left(1+\frac{\Phi_G}{c^2}\right)\cr
&&N_i = -A_i\cr
&&\gamma_{ij} = \delta_{ij} \left(1-\frac{2\Phi_G}{c^2}\right),
\eea
from which we find the intrinsic curvature of the spatial hypersurfaces up to second order 
\beq
{}^{(3)}R = \frac 4{c^2} \nabla^2 \Phi_G + \frac 2{c^4} \left[3 (\nabla\Phi_G)^2 + 8 \Phi_G \nabla^2 \Phi_G
\right]
\eeq
and, up to first order, the extrinsic curvature 
\beq
K_{ij} = \frac 1{2N}\left[\dot\gamma_{ij} - \nabla_{(i}N_{j)}\right] \approx \frac 1{2c}{\nabla_{(i}A_{j)}} = 
\frac{f_{ij}}{2c},
\eeq
where we have defined $f_{ij} = \nabla_{(i}A_{j)}$. In this approximation, the trace of the extrinsic curvature, $K = 
\eta^{ij} K_{ij}$ vanishes by virtue of the gauge condition. After some algebra, we find the gravitational action 
\beq
S_G = \beta\int dt \int d^3\vr \left(-(\nabla\Phi_G)^2 + \frac{c^2}8 f_{ij} f^{ij}\right),
\eeq
where $\beta = (8\pi G)^{-1}$. The resulting total effective action,
\beq
S = \int dt\int d^3\vr \left[\frac{i\hbar}2 \psi^* \overleftrightarrow{\nabla_t}\psi - \frac{\hbar^2}{2m} 
|\nabla\psi|^2 - \tV(|\psi|) - m\Phi_G|\psi|^2 - J\cdot A - \beta (\nabla\Phi_G)^2 + \frac{\beta c^2}8 
f_{ij} f^{ij}\right],\\
\label{GPgravaction}
\eeq
now allows us to use the non-linear Schroedinger equation to describe large scale structure, so long as $\psi(t,\vr)$ is 
interpreted as a Schroedinger {\it field}, normalized to the particle number, as mentioned earlier.

\subsection{Equations of Motion}

Extremizing the action in \eqref{GPgravaction} with respect to variations in $\psi$ and employing the gauge 
conditions gives non-linear Schroedinger equation for $\psi$
\beq
i\hbar \mathcal{D}_t \psi = -\frac{\hbar^2}{2m} \nabla^2\psi + m\Phi_G\psi + \tV'(|\psi|)\psi
\label{GPschrod}
\eeq
where $\mathcal{D}_t = \left(\nabla_t-  A^i \nabla_i\right)$ is the transport derivative, which takes into 
account frame dragging due to the rotation. Likewise, varying $A_i$ and $\Phi_G$ give
\beq
\nabla_j f^{ji} = \nabla^2 A^i = -\frac{16\pi G}{c^2}{J^i}
\label{Aeqn}
\eeq
(employing the gauge condition) and
\beq
\nabla^2 \Phi_G = 4\pi G m |\psi|^2,
\label{Phieqn}
\eeq
up to first order. These are readily solved by
\bea
&&\Phi_G(\vr) = -4\pi G m\int d^3 \vr'~ \frac{\psi^*(t,\vr')\psi(t,\vr')}{|\vr - \vr'|}\cr\cr
&&A_i(\vr) = \frac{8\pi i G\hbar}{c^2}\int d^3 \vr'~ \frac{\psi^*(t,\vr')\overleftrightarrow{\nabla'}_i\psi(t,\vr')}
{|\vr - \vr'|}
\label{gravcontr}
\eea
respectively. 

It is often useful to treat the BEC as a superfluid. We now consider the hydrodynamic analogy by employing the Madelung 
transformation \cite{m27} to rewrite equation \eqref{GPschrod}. Setting \cite{phc11,bh07,phc16}
\beq
\psi(t,\vr) = \sqrt{\nu(t,\vr)}~e^{i S(t,\vr)}
\eeq
where $\nu$ is the number density of particles and $S$ is the real action. By comparing the real and imaginary parts 
of the Schroedinger equation for $\psi$ we find 
\bea
&&\cD_t\nu + \frac\hbar m \left(\nabla \nu \cdot \nabla S + \nu \nabla^2 S\right)=0\cr\cr
&&\cD_t S + \frac\hbar{2m} (\nabla S)^2 + \frac m\hbar \Phi_G + \frac 1\hbar \tV'(\nu) + Q(\nu) = 0
\eea
where $Q$ is the ``quantum potential''
\beq
Q = -\frac\hbar{4m} \left[\frac{\nabla^2\nu}\nu - \frac 12 \left(\frac{\nabla\nu}\nu\right)^2\right]
\eeq
If we introduce the velocity field, $u = \hbar \nabla S/m$, then the first of the above equations,
\beq
\cD_t\nu + \nabla\cdot(\nu u) = 0,
\eeq
has the form of a continuity equation. The second can also be put in an interesting form: write it as
\beq
\nabla_t u - \nabla(A\cdot u) + \frac 12 \nabla (u)^2 + \nabla\Phi_G + \nabla\left(\frac 1 m 
\tV'(\nu)\right) + \frac{\hbar \nabla Q}m = 0
\eeq
and notice that $u$ is irrotational, so $\nabla_i u_j = \nabla_j u_i$ implying that $\nabla(u^2) = 2(u\cdot 
\nabla) u$. Likewise $\nabla_i(A\cdot u)= (A\cdot \nabla)u_i + u_j\nabla_i A_j$, therefore
\beq
\cD_t u_i + (u\cdot \nabla)u_i = - \frac 1{m\nu}\nabla_i P - \nabla_i \Phi_G +  u_j \nabla_i A_j - \frac\hbar 
m \nabla_i Q
\eeq
where $m\nu$ represents the mass density of the BEC and $P$ is the pressure given by
\beq
\nabla_i P =  \nu \nabla_i \tV'(\nu).
\eeq
This is the general form of the equation of state. For example, for quartic interactions, $\tV(\nu)= \frac\lambda 
4 \nu^2$, we find \cite{bh07}
\beq
P = \frac \lambda 4 \nu^2.
\eeq
Thus, not surprisingly, a repulsive interaction ($\lambda>0$) leads to a positive pressure and an attractive interaction
($\lambda<0$) to a negative pressure. The effect of frame dragging is contained in the transport derivative, $\cD_t$.
In terms of the particle density and real action, the gravitational potential and shift are given by
\bea
&&\Phi_G(\vr) = -4\pi G m\int d^3 \vr'~ \frac{\nu(t,\vr')}{|\vr - \vr'|}\cr\cr
&&A_i(\vr) = -\frac{16\pi Gm}{c^2}\int d^3 \vr'~ \frac{\nu(t,\vr')u_i(t,\vr')}
{|\vr - \vr'|}
\label{gravcontr2}
\eea
respectively.

\section{Stable Configurations}

(Meta)stable, rotating BEC configurations may be obtained by extremizing the total energy of the system, which, according to 
\eqref{GPgravaction} will be given by
\beq
H = \int d^3 \vr\left[\frac{\hbar^2}{2m} |\nabla_i\psi|^2 + V(|\psi|) + \frac m2 \Phi_G |\psi|^2 + \frac 12 J 
\cdot A\right].
\label{totalHam}
\eeq
Introducing a length parameter, $R$, a general ansatz for $\psi(\vr)$ representing a rotating condensate in spherical 
coordinates would be
\beq
\psi(t,\vr) = w\sum_{k\geq |l| = 0,1,\ldots}  F_{kl}(r/R) Y_{kl}(\theta,\phi) e^{i\mu t},
\eeq
where $w$ is a normalization, $\mu$ is the non-relativistic chemical potential associated with the BEC, $k,l$ are integers, 
$-k\leq l\leq k$, $F_{kl} (r/R)$ are real functions of $r$ and $Y_{kl}$ are the spherical harmonics. In what follows, we 
will treat $F_{kl}(r/R)$ as a variational function. Here $R$ is a variational parameter. We hold the particle number 
and the total angular momentum fixed in the variational computation.

\subsection{The Total Energy}

For simplicity, we will analyze the situation when all particles are in the $k=1=l$ eigenstate of angular momentum, 
{\it i.e.,} we take 
\beq
\psi(t,\vr) =  w F(r/R) \sin\theta e^{i\varphi} e^{i\mu t}.
\eeq
The procedure below can be extended to arbitrary angular momentum states. The normalization condition gives,
\beq
w = \sqrt{\frac{3N}{8\pi R^3C_{22}}}
\eeq
where $N$ is the number of bosons in the BEC and $C_{22} = \int_0^\infty d\xi~ \xi^2 F^2(\xi)$. To find an expression for 
$\Phi_G$, we use the expansion of the Green function in spherical harmonics with vanishing boundary conditions at the origin 
and at infinity,
\beq
\frac 1{|\vr-\vr'|} = {4\pi}\sum_{l,m} \frac 1{2l+1} \frac{r_<^l}{r_>^{l+1}} Y^*_{lm}(\theta',\varphi')
Y_{lm}(\theta,\phi)
\eeq
and find
\bea
\Phi_G(r,\theta) &=& -\frac{3G m N}{2C_{22}R}\left\{\frac 23\left[\frac Rr \int_0^{r/R} d\eta \eta^2 F^2(\eta) + 
\int_{r/R}^\infty d\eta \eta F^2(\eta)\right]\right.\cr\cr
&&\left.\hskip 1.5cm + \frac 1{15}(1-3\cos^2\theta) \left[\frac{R^3}{r^3} \int_0^{r/R} d\eta~ \eta^4 F^2(\eta) + 
\frac{r^2}{R^2}\int_{r/R}^\infty \frac{d\eta}{\eta} F^2(\eta)\right]\right\}.
\eea
Now from \eqref{gravcontr} it follows that only $A_\phi$ survives (in this stationary state), and 
\beq
A_\phi(r,\theta) = \frac{4\hbar}{mc^2}~ \Phi_G(r,\theta).
\eeq
Putting these results into \eqref{totalHam}, and taking, for simplicity, quartic interactions of arbitrary 
sign, $V(|\psi|) = \frac\lambda 4 |\psi|^4$, we find the four terms in the total energy to be
\bea
H_K(R) &=& \frac{N\hbar^2}{2mR^2 C_{22}} \int_0^\infty d\xi\left[\xi^2 F'^2(\xi) + 2 F^2(\xi)\right]\cr\cr
H_V(R) &=& \frac{3\lambda N^2}{40\pi R^3 C_{22}^2} \int d\xi~ \xi^2 F^4(\xi)\cr\cr
H_\Phi(R) &=& -\frac{Gm^2 N^2}{2RC_{22}^2}\left[\int_0^\infty d\xi~ \xi F^2(\xi) \int_0^\xi d\eta~ \eta^2 F^2(\eta)
+ \int_0^\infty d\xi~ \xi^2 F^2(\xi) \int_\xi^\infty d\eta~ \eta F^2(\eta)\right.\cr\cr
&& \hskip 1cm \left.+ \frac 1{25}\int_0^\infty \frac{d\xi}\xi F^2(\xi) \int_0^\xi d\eta~ \eta^4 F^2(\eta) + 
\frac 1{25}\int_0^\infty d\xi~ \xi^4 F^2(\xi) \int_\xi^\infty \frac{d\eta}\eta F^2(\eta)\right]\cr\cr
H_A(R) &=& \frac{3\hbar^2 G N^2}{c^2 C_{22}^2 R^3}\left[\int_0^\infty \frac{d\xi}\xi F^2(\xi) \int_0^\xi 
d\eta~ \eta^2 F^2(\eta) + \int_0^\infty d\xi F^2(\xi) \int_\xi^\infty d\eta~\eta F^2(\eta)\right]
\eea
where $\xi=r/R$ is a dimensionless variable. The second term in the integral in the expression for the kinetic energy, 
$H_K$, represents the contribution of the azimuthal motion of the condensate. $H_\Phi$ and $H_A$ are the contributions 
of the gravitational field. These expressions may be put into simpler form in terms of the coefficients
\bea
&&C_{mn} = \int_0^\infty d\xi~\xi^m F^n(\xi)\cr\cr
&&B_{mn} = \int_0^\infty d\xi~\xi^m F'^n(\xi)\cr\cr
&&D_{mn} = \int_0^\infty d\xi~ \xi^m F^2(\xi) \int_0^\xi d\eta~ \eta^n F^2(\eta)\cr\cr
&&A_{mn} = \int_0^\infty d\xi~ \xi^m F^2(\xi) \int_\xi^\infty d\eta~ \eta^n F^2(\eta).
\label{coeffs}
\eea
We find
\bea
H(R) &=& H_\Phi(R) + H_K(R) +  H_V(R) + H_A(R)\cr\cr
&=&  -\frac{Gm^2 N^2}{2RC_{22}^2}\left(D_{12}+A_{21}+\frac 1{25}(D_{-14}+A_{4,-1})\right) + \frac{N\hbar^2(B_{22}+2C_{02})}
{2mR^2 C_{22}}\cr\cr
&&\hskip 4cm + \frac{3\lambda N^2C_{24}}{40\pi R^3 C_{22}^2} + \frac{3 G\hbar^2 N^2 }{c^2R^3C_{22}^2}\left(D_{-12}+A_{01}
\right).
\eea
$H(R)$ does not include the rest mass energy, $Nmc^2$, of the condensate and the total energy can be written as $E(R) \approx 
Nmc^2 + H(R)$. If $H_B(R)$ is the binding energy per particle, then $E(R) = N|mc^2-H_B(R)|$. We can therefore identify 
$|H_B(R)| = |H(R)|/N$. In fact, using previous results \cite{esvw14}, let us define the dimensionless parameters $\rho$ and $n$ by
\beq
R = \frac{M_p}m \sqrt{\frac{|\lambda|}{\hbar c}}~ \rho,~~ N = \frac{nM_p}{m^2} \left(\frac\hbar c\right)^{3/2} \frac c
{\sqrt{|\lambda|}},
\label{dimensionfree}
\eeq
where $M_p = \sqrt{\hbar c/G}$ is the Planck mass. Then the binding energy per unit mass may be given as,
\beq
\mathcal{H}(\rho) = \frac H{mN} = \frac{\hbar^3 c}{|\lambda| M_p^2}\left[\frac A\rho + \frac B{\rho^2} + \frac C{\rho^3}\right],
\label{enpermass}
\eeq
where
\bea
&&A = -\frac n{2C_{22}^2}\left(D_{12}+A_{21}+\frac 1{25}(D_{-14}+A_{4,-1})\right)\cr
&&B = \frac 12 \left(\frac{B_{22}+2C_{02}}{C_{22}}\right)\cr
&&C = n\left(\sgn(\lambda) \frac{3C_{24}}{40\pi C_{22}^2} + \frac{3 \hbar^3}{|\lambda| c M_p^2 C_{22}^2}(D_{-12}+A_{01})\right).
\label{constants1}
\eea
The coefficient $A$ characterizes the contribution of the gravitational potential energy to the total energy and is negative and 
$B$ represents the contribution of the kinetic energy of the bosons. The effects of the rotation are contained in a contribution 
to the kinetic energy through the coefficient $C_{02}$ in $B$ and in the last term in the expression for $C$, which combines the
contributions of the scalar potential and frame dragging. This term is proportional to $3n/b^2$, where $b$ is a dimensionless 
parameter, $b^2=|\lambda| c M_p^2/\hbar^3$, characterizing the strength of the self interactions.

\subsection{Minimum Energy Configuration}

The extrema of $\cH$ in \eqref{enpermass} are readily found to be given by
\beq
\rhoeq = \frac{B}{|A|} \mp \sqrt{\frac{B^2}{|A|^2} + \frac{3C}{|A|}},
\eeq
assuming it is real, {\it i.e.,} provided that $B^2 > 3|A||C|$ when $C<0$. In what follows, we set 
\bea
A &=& a n\cr
C &=& qn\left(\frac d{qb^2} + \sgn(\lambda)\right)
\label{constants12}
\eea
where the parameters $a$, $q$ and $d$ are obtained from the constants defined in \eqref{constants1}, {\it viz.,} 
\bea
&&a = -\frac 1{2C_{22}^2}\left(D_{12}+A_{21}+\frac 1{25}(D_{-14}+A_{4,-1})\right)\cr\cr
&&q = \frac{3C_{24}}{40\pi C_{22}^2}\cr\cr
&&d = \frac 3{C_{22}^2} \left(D_{-12}+A_{01}\right).
\label{constants3}
\eea
The first , $a$, characterizes the strength of the gravitational interaction, the second, $q$, characterizes the strength of 
the self interaction and the last, $d$, the effect of the frame dragging. In terms of these, the dimensionless equilibrium 
radius is
\beq
\rhoeq = \frac B{|a|n}\left[1 + \sqrt{1 + \frac{3|a|q}{B^2}\left(\frac d{qb^2} + \sgn(\lambda)\right)n^2}\right].
\label{eqrad}
\eeq
For it to exist, we must require the quantity under the radical to be non-negative. If the kinetic energy of the bosons
is ignored, as in the Thomas-Fermi approximation, $B$ approaches zero and the equilibrium radius approaches a constant
independent of the number of particles. In the absence of rotation ($d=0$) this equilibrium is possible only for 
repulsive self interactions. If one does not ignore the kinetic energy of the bosons then the equilibrium radius 
decreases with increasing $n$.

When $C>0$ an absolute minimum of the energy exists, as the interactions together with the rotations are able to stabilize the 
condensate. When $C<0$ only attractive interactions are permitted and the energy of the system is unbounded from below, but we 
have found a local minimum in \eqref{eqrad}, not an absolute minimum, of the energy. This local minimum exists only when the 
number of particles is below a certain critical value, $\nc$, which will be determined below. It is important to discuss 
the validity of this solution.  

There are three criteria that it must satisfy in order to be self consistent, the first two arising 
from the GEM approximation. First, the de Broglie wavelength of the bosons should be much larger than their Compton wavelength 
to ignore special relativistic corrections. Second, the size of the condensate at the minimum has to be much  greater than the 
Schwarzschild radius of the corresponding mass, to justify not using general relativistic corrections. The impact of these two 
conditions on the range of the parameters is determined below and the conditions are satisfied by the examples we give. When 
the energy is unbounded from below (in the case $C<0$, $\lambda<0$) the local minimum we have found could be unstable. This can 
be avoided if additional stabilizing terms exist in the effective potential (as is the case for QCD axions, where the effective 
potential is given by the Bessel function $J_0$ \cite{esvw14}). Irrespective of whether there is a stabilization term,  the 
configuration at  the local minimum could be a viable state,  provided it has a lifetime greater than the age of the universe. 
This would occur if the probability of tunneling out of  it is sufficiently small. The lifetime of metastable BECs formed by 
bosonic atoms with attractive interactions, confined by a harmonic trap, was estimated using instanton methods in 
\cite{st97,frar99}. One can use similar techniques to compute the lifetimes of gravitationally bound BEC systems against tunneling 
out of the metastable local minimum state. The result is that the tunneling rate behaves approximately as $\sim \exp(-2NJ/\hbar)$, 
where $N$ is the number of particles and $J$ is the WKB expression for the exponent of the tunneling rate. In the case of an 
astrophysical BEC, the number of particles is very large ($N \sim 10^{58}$ -- $10^{100}$) and the WKB expression, $J$, is not 
vanishingly small unless the fraction $f=n/\nc$ is commensurately close to unity. Thus so long as $f<1$ the lifetime of the 
metastable state grows exponentially with the particle number.

\subsubsection{$C>0$}

If $C>0$ there is an absolute minimum of the energy and either $\lambda>0$ or 
\beq
\lambda<0~~ \text{and}~~ b <  \sqrt{\frac dq}.
\label{ubb}
\eeq
For attractive self-interactions ($\lambda<0$), rotation provides a stabilization mechanism. In the absence 
of rotation ($d=0$), $C>0$ is possible only for repulsive interactions. 

So long as $C>0$, the number of particles, $n$, and the interaction strength, $b$, are limited only by the GEM
approximation. As mentioned, we obtain a criterion for the validity of the non-relativistic approximation by requiring that 
the de Broglie wavelength of the bosons, $\lambda_\text{dB}$, is much larger than the Compton wavelength, $\lambda_\text{C}$. 
As the de Broglie wavelength is roughly the typical size of the condensate, which is on the order of the scale factor, $R$, 
for a BEC and, from \eqref{dimensionfree}, $R_\text{eq} \sim b \rhoeq \lambda_\text{C}$, it follows that the condition for 
the non-relativistic approximation to hold is 
\beq
b\rhoeq\gg 1. 
\label{nrapprox}
\eeq
To justify not including higher order general relativistic effects one also wants a stable configuration whose equilibrium 
radius is much larger than the Schwarzschild radius, $\rho_S = 2n/b^2$, so requiring that $\rho_S \ll \rhoeq$, we find
\beq
n \ll \frac{b}2\sqrt{\frac{3q}{|a|}}\sqrt{\frac{4B+3d}{3q}+b^2\sgn(\lambda)}~~ \stackrel{\text{def}}{=}~~ n_S,
\label{Cg0rest}
\eeq
which places an upper limit on the particle number for any given value of $b$.

It will be seen from \eqref{nrapprox} and \eqref{Cg0rest} that, for attractive self-interactions, the weak gravity approximation 
holds only for condensates with small $n$ because $b$ is bounded from above according to \eqref{ubb}. On the contrary, there is 
no such restriction on $n$ for repulsive interactions as $b$ can be arbitrarily large \cite{csw86}. As mentioned, in this case 
\eqref{eqrad} represents an absolute minimum of the energy. Increasing the particle number, $n$, also increases the strength of 
the gravitational attraction, therefore the radius of the system decreases with increasing $n$. In the limit of large $n$ 
(repulsive self interactions) the system approaches the equilibrium radius,
\beq
\rho_\infty = \sqrt{\frac{3q}{|a|} \left(\frac d{qb^2} + 1\right)}.
\eeq

\subsubsection{$C<0$}

If $C<0$ the energy is not bounded from below, but there is a local minimum of the energy. This case is only possible when
\beq
\lambda<0~~ \text{and}~~ b > \sqrt{\frac dq}.
\eeq
The condition for the validity of the non-relativistic approximation is \eqref{nrapprox}, the same as in the case $C>0$ above.
As $b$ can be arbitrarily large, this condition can always be verified. The energy is not bounded from below, but a local minimum 
exists provided that there is an an upper limit on the number $n$ 
\beq
n \leq \frac B{\sqrt{3|a|q}} \left(1-\frac d{qb^2}\right)^{-1/2}~ \stackrel{\text{def}}{=}~ \nc.
\label{ncritical}
\eeq
This therefore defines a critical number of particles above which there is no metastable configuration. The local 
equilibrium radius, which can now be written as 
\beq
\rhoeq = \frac B{|a|n}\left[1 + \sqrt{1 - \frac{n^2}{\nc^2}}\right],
\eeq
continues to decrease with increasing $n$, so the smallest metastable condensate size occurs for $n=\nc$. The 
existence of a minimum size (maximum number of particles) for the case of negative self-interactions can be 
understood as follows: for an equilibrium solution it is necessary for the inward gravitational and self 
interaction pressures to be balanced by the outward quantum pressure. This condition was used in \cite{ds16}
to estimate the mass and radius of a non-rotating axion drop. They found an upper bound on the drop mass (as
previously determined by Chavanis \cite{phc11}) as well as the virial relation between the radius and mass 
of an object supported against gravity by pressure. 

Finally, to consistently ignore general relativistic corrections, we ask that $\nc$ is small enough so that the 
equilibrium radius remains much larger than the Schwarzschild radius. This gives
\beq
\nc \ll b\sqrt{\frac{B}{2|a|}}~ \stackrel{\text{def}}{=}~ n_S
\label{Cl0rest}
\eeq
and $\rho_c = B/(|a|\nc)$ is the critical radius.

\subsection{Single Vortex Ansatz}

We can get a rough idea of the size of the rotating condensates by taking a variational approach \cite{phc11,ps08,
rds12,elsw16}. This approach is known to be in good agreement with more precise (numerical) analyses \cite{kaup68,
rb69,bb11,esvw14,dc11,eknw15,ps08} and requires some ansatz for the function $F(\xi)$. We ask for a trial 
wavefunction that behaves as a vortex of width $R$ and that is continuous everywhere. As a trial wave function, 
it is not required to be a solution of the equations of motion. Continuity at the origin suggests that, as $\xi$ 
approaches zero, $F(\xi)$ should behave as $\xi^l$, where $l\geq 1$. We therefore take 
\beq
F(\xi) = \xi e^{-\xi^2/2},
\label{Fxi}
\eeq
so that the wavefunction in cylindrical coordinates has the form
\beq
\psi(\zeta,\phi) = \zeta e^{-(\zeta^2+z^2)/R^2} e^{i\phi}
\eeq
where $\zeta$ is the cylindrical radius. The gravitational potential becomes
\bea
\Phi_G(r,\theta) &=& -\frac{G m N}r \left[\left(1+\frac{R^2}{4r^2}(1 -3\cos^2\theta)\right)\erf(r/R)\right.\cr\cr
&&\left. \hskip 2cm + \frac{r e^{-r^2/R^2}}{\sqrt{\pi} R}\left\{\left(\frac 12 + \frac{3R^2}{4r^2}\right)\cos 2\theta 
- \frac 12 + \frac{R^2}{4r^2}\right\}\right]
\label{gravitpot}
\eea
and we find
\bea
a &=& -\frac{23}{30\sqrt{2\pi}},\cr
B &=& \frac{5}4,\cr
q &=& \frac 1{16\sqrt{2}\pi^{3/2}},\cr
d &=& \frac 53 \sqrt{\frac 2\pi}.
\label{constants2}
\eea
applying \eqref{constants1} and \eqref{constants3}
\begin{figure}
\begin{minipage}[t]{8cm}
\includegraphics[scale=0.6]{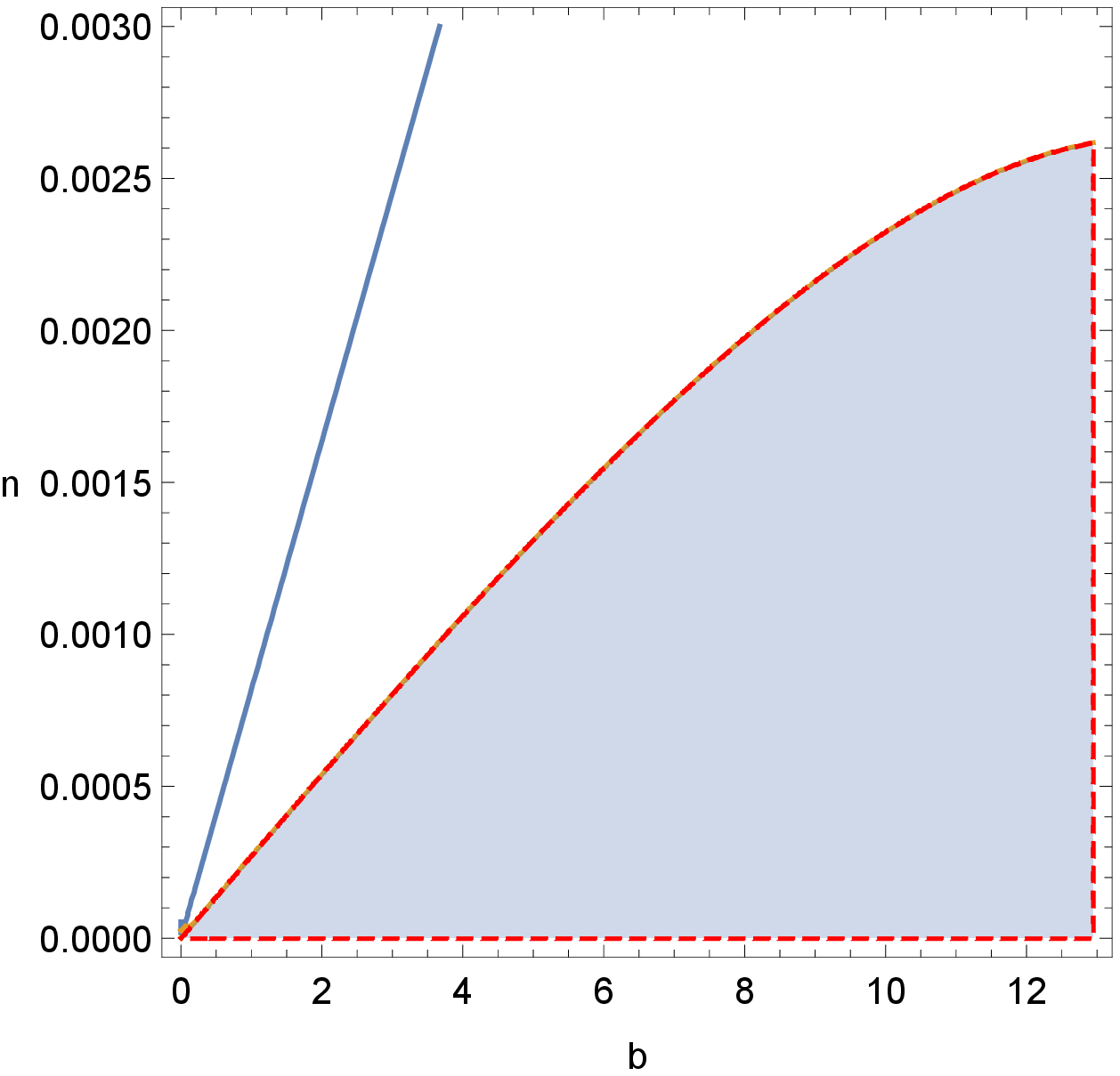}
\caption{The shaded region represents the possible values of the parameters $n$ and $b$ in \eqref{spcase} for 
which $b\rhoeq \sim n_S/n > 10^4$ for the case $C>0$ and $\lambda <0$. Here $b$ is required to be $<12.9$.}
\label{paramregionA}
\end{minipage}
\hfill
\begin{minipage}[t]{8cm}
\includegraphics[scale=0.6]{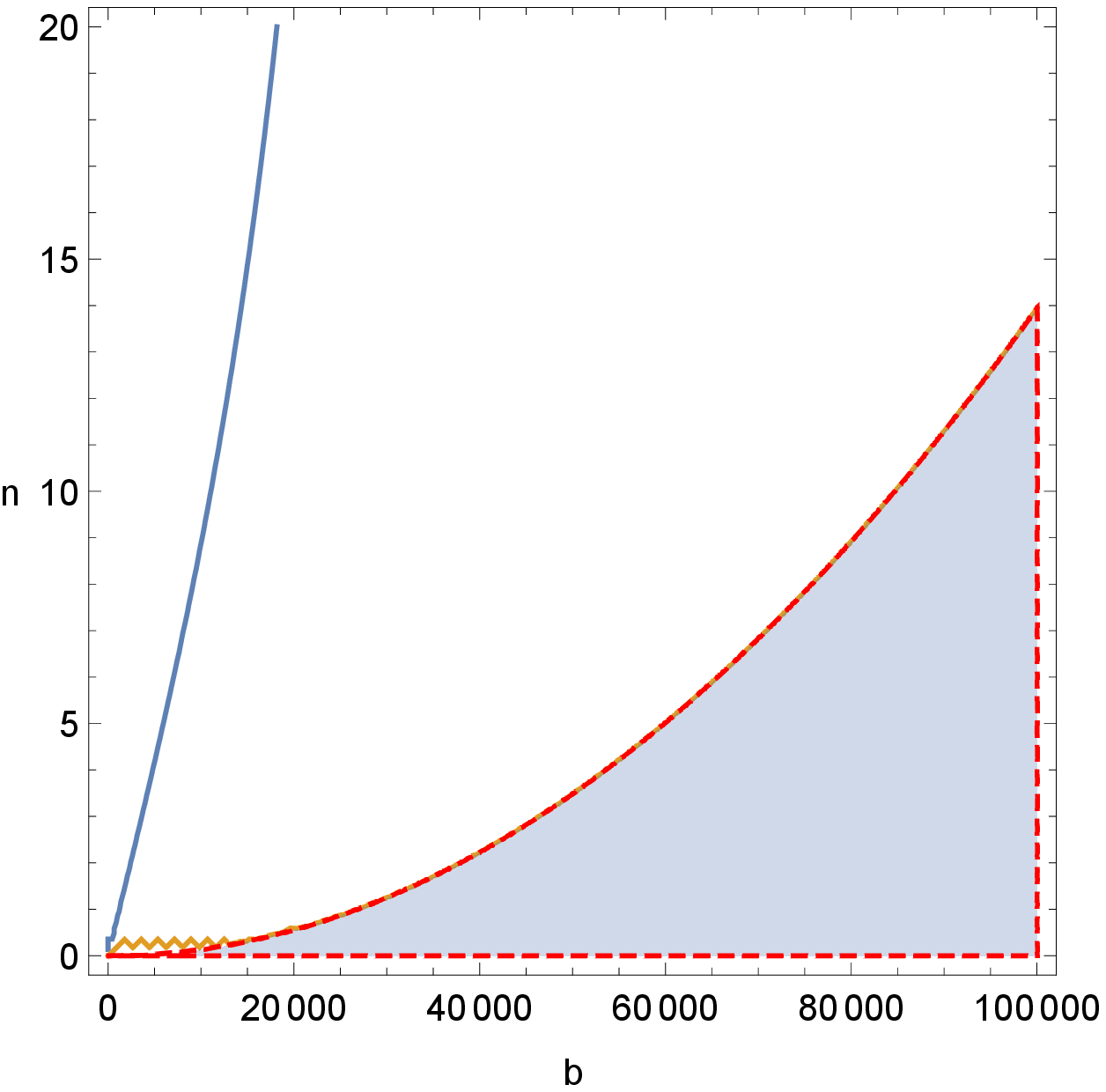}
\caption{The shaded region represents the possible values of the parameters $n$ and $b$ in \eqref{spcase} for 
which $b\rhoeq > 10^4$ and $n_S/n > 10^8$ for the case $C>0$ and $\lambda >0$. Here there is no limit on $b$.}
\label{paramregionB}
\end{minipage}
\end{figure}

When $C>0$, the conditions for the validity of the GEM approximation are \eqref{nrapprox} and \eqref{Cg0rest},
\bea
&&b\rhoeq = \frac{4.09 b}n\left[1+\sqrt{1+4.66 \times 10^{-3}\left(\frac{168}{b^2}+\sgn(\lambda)\right)n^2}\right]
\gg 1\cr\cr
&&\frac{n_S}n = \frac{0.140b}n\sqrt{378 +b^2\sgn(\lambda)} \gg 1
\label{spcase}
\eea
These conditions can be met by small condensates with attractive interactions and a suitably limited value of $b$
($b< 12.9$). For repulsive interactions there is no upper limit on $b$ and large condensates are possible. The shaded 
regions in figures \ref{paramregionA} and \ref{paramregionB} represent the portion of the parameter space satisfying these
conditions for the case of attractive interactions (figure \ref{paramregionA}) and repulsive interactions (figure 
\ref{paramregionB}). In each case, the trustworthy portion of the parameter space is limited by the second inequality 
in \eqref{spcase}. 

When $C<0$ only a metastable state may exist. The critical (maximum possible) number of bosons is 
\beq
\nc = 14.6\left(1-\frac{168}{b^2}\right)^{-1/2}.
\eeq
and the minimum possible equilibrium radius is given by the limit as $n\rightarrow \nc$, 
\beq
\rho_c = \rhoeq(n\rightarrow \nc) = 0.279 \sqrt{1 - \frac{168}{b^2}}.
\eeq
The two conditions \eqref{nrapprox} and \eqref{Cl0rest} for the validity of the GEM approximation can therefore be 
stated as
\bea
&&b\rhoc\gg 1\cr\cr 
&&\frac{n_S}{\nc} = \frac{1.43 b}\nc \gg 1.
\eea
As there is no upper limit on $b$, they are readily met so long as $b$ is sufficiently large. In addition, to ensure a 
large lifetime, the actual mass of the BEC cannot be too close to its maximum mass, {\it i.e.,} the fraction 
$f=n/\nc$ should not be too close to unity.

The parameters $\nc$ and $\rhoc$ of the BEC cloud are uniquely determined by the size of the self-interaction, $b$, 
so the length scale, $R$, and the total mass, $M=mN$ depend only on the strength of the self-interaction and the particle 
mass. 

\subsection{Asteroid vs. Galaxy Size Halos}

As an example, we consider the condensates formed by particles of mass and interaction strength typical of 
the QCD axion, $m \sim 10^{-5}$ ev and $b \sim 2 \times 10^{7}$, where the axion decay constant is taken to 
be roughly $f_a/M_p \sim 5\times 10^{-8} \left(\frac c\hbar\right)^{3/2}$. QCD axions have attractive interactions and 
the size of $b$ indicates that we are considering the case $C<0$. From \eqref{dimensionfree} we find that 
\beq
N_c = \frac \nc b\left(\frac{M_p}{m}\right)^2~ \approx 1.10 \times 10^{60},
\eeq
with a total mass of $M_c = mN_c = 1.95 \times 10^{19}$ kg. We also find the critical length scale associated 
with the ball
\beq
R_c = b\rho_c \left(\frac \hbar{mc}\right)~ \approx 5.6 \times 10^6 \frac\hbar{mc}
\eeq
The Compton wavelength of the $10^{-5}$ eV boson is about 0.02 m, which gives the radius as about $R_c 
\approx 110$ km. The radius inside of which about 99.9\% of the matter is confined, denoted by $R_{99}$, is 
roughly 3.5 times this radius, so we find $R_{99} \approx 385$ km.

\begin{figure}
\begin{center}
\includegraphics[scale=0.8]{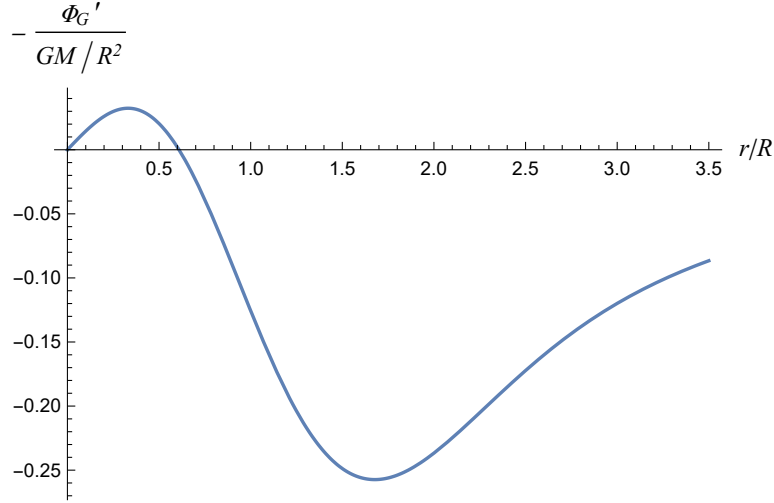}
\caption{Gravitational acceleration in the equatorial plane due to the Vortex BEC}
\label{GravForce}
\end{center}
\end{figure}

Continuing with attractive self-interactions and assuming that $b\gg 10^2$, \eqref{ncritical} suggests that we 
take the maximum (critical) size to be approximately independent of $b$. Then $\rho_c \sim 1/\nc$ is also approximately 
independent of $b$ and the dependence of $M_c$ and $R_c$ on $m$ and $b$ is simple. Using \eqref{dimensionfree} one 
finds that the particle mass and interaction strength required to produce a desired value of $R_c$ and $M_c$ are given by
\beq
m \approx \sqrt{\frac{B\hbar M_p^2}{|a|c M_c R_c}},~~ b \approx \sqrt{\frac{B c M_p^2 R_c}{3q\hbar M_c}}.
\eeq
Taking  $R_{99} \sim 50$ kpc, and a mass of roughly three times that of the visible galaxy, $M_c \approx 10^{42}$ kg, 
we find $mc^2 \sim 10^{-24}$ eV and $b \sim 10^4$. The condensate continues to be non-relativistic, satisfying the 
condition $b\rho_c \gg 1$, with a total angular momentum of $L = N\hbar \approx 10^{67}$ J$\cdot$s, which is 
comparable to that of the luminous matter \cite{cabe15}. For the vortex wavefunction in \eqref{Fxi}, however, the 
gravitational force in the equatorial plane, $F = -\nabla \left.\Phi_G\right|_{\theta=\pi/2}$, is outward directed up 
to a distance of about $0.609 R$ (see figure (\ref{GravForce})). This can be attributed to the shape of the density 
profile which vanishes at the center and increases until about $0.707 R$ as one moves outward in the equatorial plane 
before falling off, all the while decaying exponentially perpendicular to the equatorial plane. For stable orbits 
to exist within this distance from the center, the region must be dominated by ordinary (baryonic) matter. This can be 
used to set the scale for the wavefunction. For example, there is good evidence via near-infrared and optical photometry 
\cite{ipb15a,pi15,ipb15c} to suggest that the Milky Way is dominated by ordinary baryonic matter up to about 6-8 kpc 
from the center, which is roughly the location of our solar system. This implies that $R_c \approx 10-13$ kpc, which 
gives $R_{99} \approx 35-50$ kpc. On the other hand, for $r\gg R$ the gravitational force obeys the usual inverse 
square law. 

A straightforward analysis of circular orbits within the BEC on the equatorial plane at $r>0.609R$ shows that the 
tangential speed may be given as the sum of two contributions,
\beq
v^\phi = \pm \sqrt{r\Phi_G'} + \frac{\hbar\Phi_G}{mc^2 r}.
\eeq
where the prime refers to a derivative with respect to $r$. The first is due to the gravitational force, the second 
represents the effect of frame dragging. The latter is, however, negligible compared with the contribution due to the 
gravitational force. In the interior, $r<0.609R$, the expression for the tangential speeds will be the same, but the 
first term will be significantly modified by the baryonic matter, which we assume dominates. However, the contribution 
from the BEC cloud due to frame dragging persists and may get significant near the center as it depends only on the 
depth of the gravitational potential well and not its gradient.

\section{Vortex Oscillations}

To describe the approximate dynamics of an imploding BEC, we employ a collective coordinate description in terms of 
the condensate radius \cite{ps08,st97,ul98,frar99}, taking
\beq
\psi(t,\vr) = \sqrt{\frac{N}{\pi^{3/2} R(t)^3}}~ \frac r{R(t)} e^{-\frac{r^2}{2R(t)^2}}\sin\theta~ 
e^{\frac{imr^2}{2\hbar}\Gamma(t) + i\phi}
\label{timedepansatz}
\eeq
where $R(t)$ and $\Gamma(t)$ represent two independent variables characterizing the dynamics of the imploding cloud.
Instead of proceeding via the Madelung equations, it is more expedient to obtain the action of the condensate
\eqref{GPgravaction} as a functional of these two variables and vary this action to obtain equations of motion for 
$R(t)$ and $\Gamma(t)$.

Applying \eqref{gravcontr2} and integrating by parts, the action \eqref{GPgravaction} can be expressed as 
\bea
S &=& \int dt\int d^3\vr \left[\frac{i\hbar}2 \psi^* \overleftrightarrow{\nabla_t}\psi - \frac{\hbar^2}{2m} 
|\nabla\psi|^2 - \tV(|\psi|) - \frac 12 m\Phi_G|\psi|^2 - \frac 12 J\cdot A\right]\cr\cr
&=& -\int dt \left[N \left(\frac 54 m(R^2 \dot \Gamma + \Gamma^2 R^2)+\frac{5\hbar^2}{4mR^2}\right) \right.\cr
&&\hskip 1cm \left. + N^2 \left(\frac 1{R^3}\left\{\frac{5\sqrt{2}G\hbar^2}{3c^2}+\frac\lambda{16\sqrt{2}\pi^{3/2}}\right\}
- \frac{23Gm^2}{30\sqrt{2\pi} R} + \frac{288Gm^2\Gamma^2R}{25c^2\sqrt{\pi}} - \frac{499Gm^2\Gamma^2R}{50c^2\sqrt{2\pi}}\right)\right]
\cr\cr
&&
\eea
The time dependence in \eqref{timedepansatz} implies that the radial component of the shift, $A_r$, is no longer vanishing, 
as it was in the stationary case, and contributes to the equations of motion. Varying with respect to $\Gamma(t)$ we find
\beq
\Gamma(t) = \frac{125c^2\sqrt{\pi}\dot R(t)}{(1152-499\sqrt{2})GmN + 125\sqrt{\pi}c^2R(t)}.
\eeq
It is better to work with the dimensionless variables defined in \eqref{dimensionfree}, in terms of which $\Gamma(t)$ may be 
expressed as 
\beq
\Gamma(t) = \frac{\dot \rho(t)}{\rho(t)+ \frac{1152-499\sqrt{2}}{125\sqrt{\pi}}\frac n{b^2}}.
\eeq
Inserting this into the equation of motion for the condensate, $\partial L/\partial R(t) = 0$, yields,
\beq
\ddot \rho + \frac{\dot \rho^2}{2\rho\left(1+ \mu\rho\right)} = F(\rho)
\label{eomrho}
\eeq
where 
\beq
F(\rho) = -\frac{m^2c^4}{h^2}\left[\frac{P_1}{\rho^2} + \frac{P_2}{\rho^3} + \frac{P_3}{\rho^4} + \frac{P_4}{\rho^5}\right]
\eeq
and $\mu~, P_1,~ P_2,~ P_3$ and $P_4$ are the dimensionless coefficients 
\bea
&&\mu = \frac{125\sqrt{\pi}}{1152-499\sqrt{2}}\frac{b^2}n\cr\cr
&&P_1 = \frac{23 n}{75\sqrt{2\pi}~b^4}\cr\cr
&&P_2 = \frac{23(576\sqrt{2}-499)n^2-9375\pi b^2}{9375\pi b^6}\cr\cr
&&P_3 = \frac{3n(-\sgn(\lambda) 25\sqrt{2} b^2 + 16(-384 + 83\sqrt{2})\pi)}{2000\pi^{3/2}b^6}\cr\cr
&&P_4 = - \frac{(\sgn(\lambda) 3b^2+160\pi) (576\sqrt{2}-499)n^2}{5000\pi^2 b^8}.
\eea
The first integral of the motion is easily given as
\beq
\frac 12 \left(\frac{\mu\rho}{1+\mu\rho}\right)\dot \rho^2 - \mu\int^\rho d\rho'~ \frac{\rho' F(\rho')}{1+\mu\rho'}  = E,
\label{eom2}
\eeq
so we may identify $E$ with the total energy of the system,
\beq
m_\text{eff} (\rho) = \left(\frac{\mu\rho}{1+\mu\rho}\right)
\eeq
with its effective mass, and 
\beq
V_\text{eff}(\rho) = - \mu\int^\rho d\rho'~ \frac{\rho' F(\rho')}{1+\mu \rho'} 
\eeq
with the effective potential energy. Direct integration reveals that 
\beq
V_\text{eff} = \frac{2 m^2 c^4}{5 \hbar^2 b^4} \left[\frac A\rho + \frac B{\rho^2} + \frac C{\rho^3}\right]
\eeq
where $A,~ B$ and $C$ are given \eqref{constants1}, \eqref{constants12} and \eqref{constants2}, which confirms that equilibrium 
($\dot\rho = 0$) is achieved according to the conditions laid out in section III C. We will now consider the dynamical collapse 
of a rotating BEC that begins at a radius larger than the equilibrium radius. 

The equilibrium energy, $E_\text{eq}$, can be determined from the equilibrium radius, $E_\text{eq} = V(\rhoeq)$. In terms of 
the rescaled time and energy,
\beq
\tau = \left(\frac{2 m^2 c^4}{5 \hbar^2 b^4}\right)^{1/2}t,~~ \cE = \left(\frac{5 \hbar^2 b^4}{2 m^2 c^4}\right) E,
\eeq
the rescaled form of \eqref{eom2},
\beq
\frac 12 m_\text{eff}(\rho)\left(\frac{d\rho}{d\tau}\right)^2 + \oV_\text{eff}(\rho)  = \cE,
\eeq
has the general solution
\beq
\tau-\tau_0 = - \int^\rho d\rho' \sqrt{\frac{m_\text{eff}(\rho')}{2[\cE-\oV_\text{eff}(\rho')]}},
\eeq
where 
\beqn
\oV_\text{eff}(\rho) = \frac A\rho + \frac B{\rho^2} + \frac C{\rho^3}
\eeq
A solution beginning with a total energy $\cE>\oV_\text{eff}(\rhoeq)$ will collapse and will oscillate about the local minimum of 
$\oV_\text{eff}(\rho)$, provided that $\cE<\oV_\text{max}$ (in the case of attractive self-interactions). 

\begin{figure}
\begin{minipage}[t]{8cm}
\includegraphics[scale=0.6]{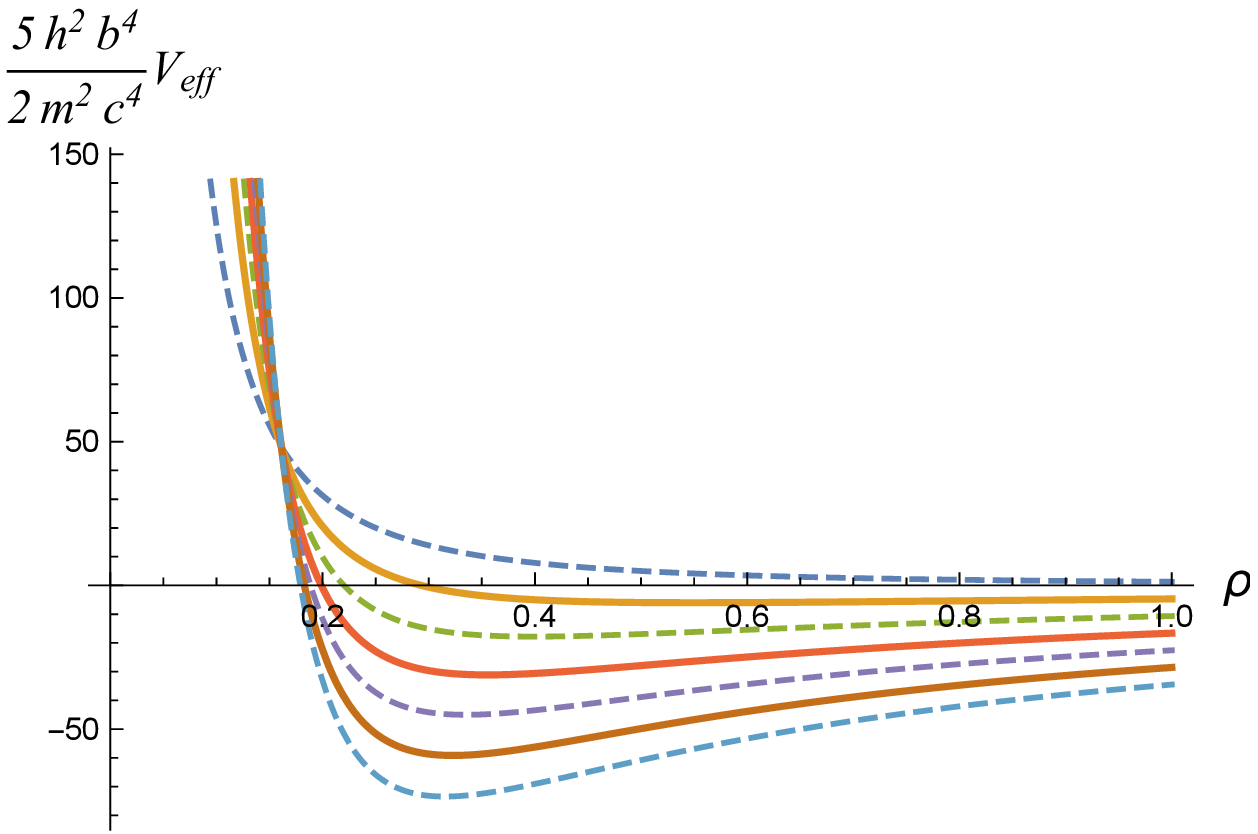}
\caption{The effective potential for a rotating BEC with repulsive interactions as a function of radius, $\rho$, for various values 
of $n$. We have taken $b=10^4$. The bottom most curve represents $n = 140$, the uppermost $n=20$. All curves admit global minima.}
\label{EffPotpositive}
\end{minipage}
\hfill
\begin{minipage}[t]{8cm}
\includegraphics[scale=0.6]{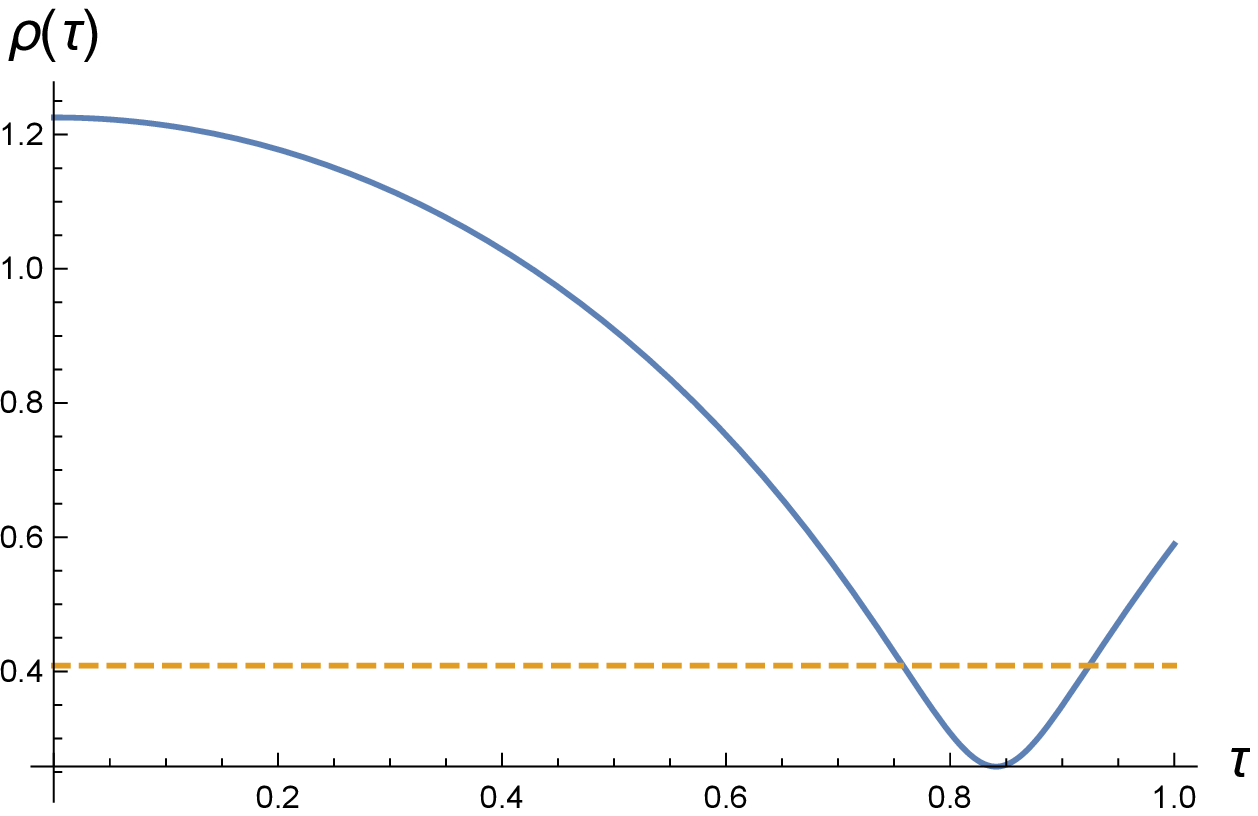}
\caption{Collapse and rebounce for a rotating BEC with repulsive interactions begining at $\rho(0)=3\rhoeq$ with zero initial 
velocity. We have taken $n=37.5$ and $b=10^4$. The dashed line is the equilibrium value of $\rho$.}
\label{ColRebpositive}
\end{minipage}
\end{figure}

BEC clouds with repulsive self-interactions always admit a global minimum of the energy functional. This is because the repulsive 
interactions in combination with the quantum pressure ensures that the potential energy grows without bound as $\rho\rightarrow 0$
(see figure (\ref{EffPotpositive})). There is therefore no limit to the size of the condensate apart from the requirement that 
the equilibrium radius is larger than the Schwarzschild radius. However, if we consider a halo of mass $M\approx 10^{42}$ 
kg made of condensed bosons of mass energy $mc^2 \approx 10^{-24}$ ev and $b \approx 10^4$, we 
find $n\approx 37.5$ and $\rhoeq \approx 0.409$ ($b\rhoeq \sim 4\times 10^3\gg 1$). Figure (\ref{ColRebpositive}) represents a 
numerical integration of \eqref{eomrho} with these conditions if the halo is assumed to begin with zero initial velocity at $\rho(0) 
= 3\rhoeq$, and shows the collapse rebounce and subsequent oscillation of the cloud about equilibrium. The cloud collapses in 
$\tau \approx 0.76$ or approximately 2.5 by.

\begin{figure}
\begin{minipage}[t]{8cm}
\includegraphics[scale=0.6]{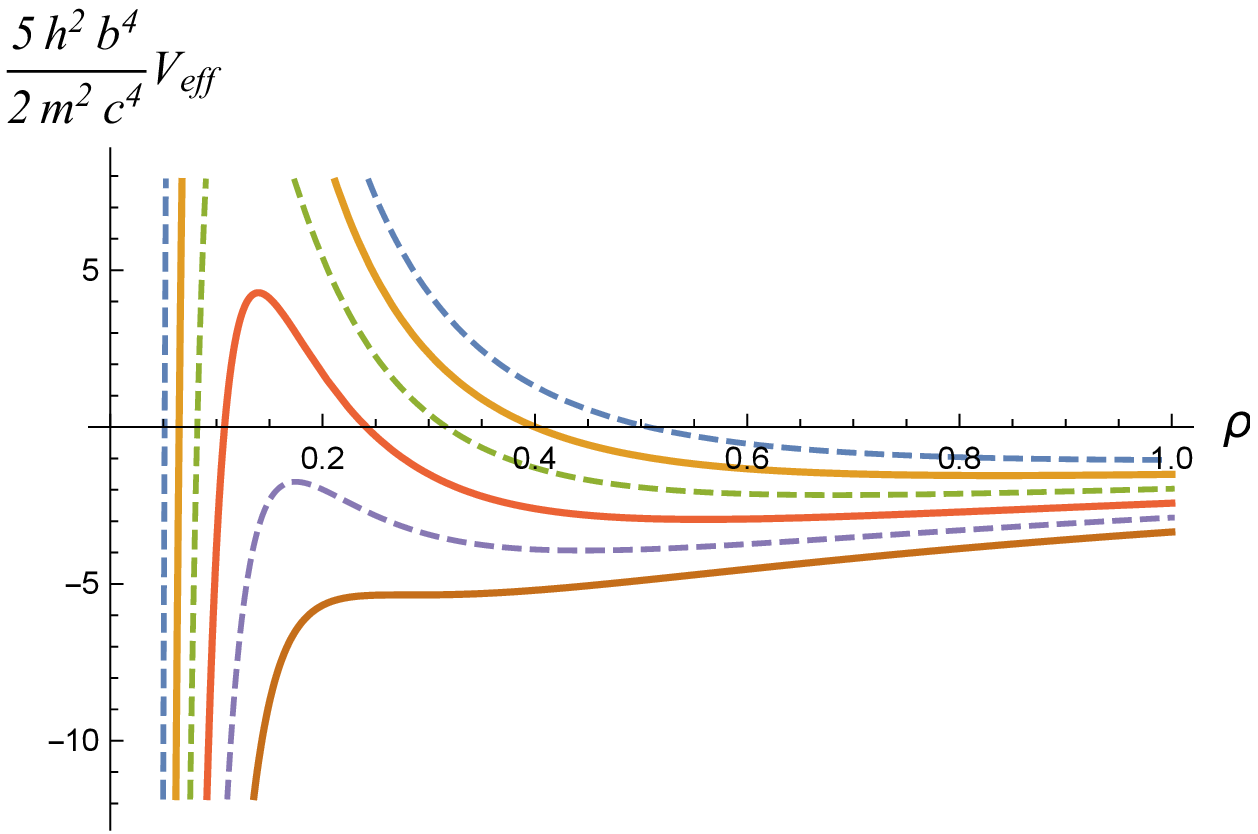}
\caption{The effective potential for a rotating BEC with attractive interactions as a function of radius, $\rho$, 
for various values of $f=n/\nc$. We have taken $b=10^4$. The bottom most curve represents $f = 1$, the uppermost $f=0.5$. 
The local minimum is shallow as $f\rightarrow 1$ and becomes deeper as $f$ decreases.}
\label{EffPotnegative}
\end{minipage}
\hfill
\begin{minipage}[t]{8cm}
\includegraphics[scale=0.6]{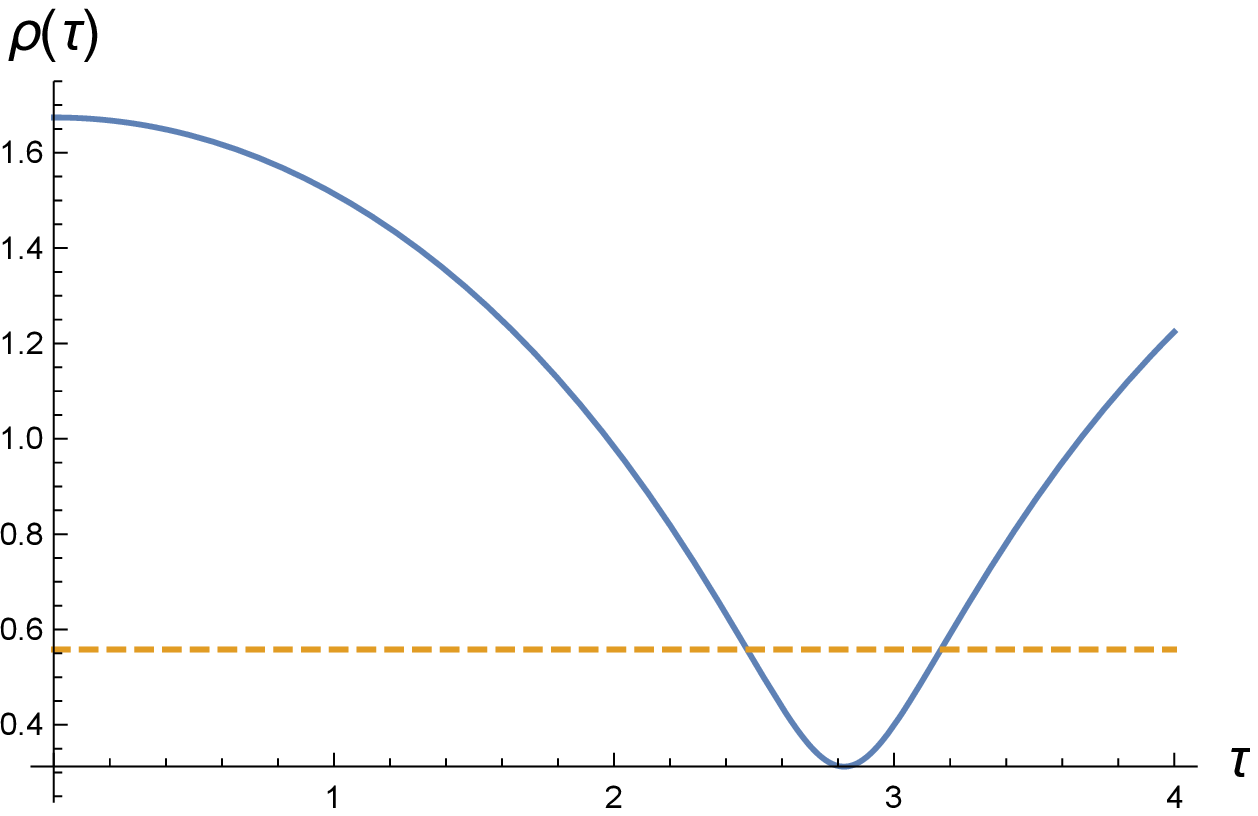}
\caption{Collapse and rebounce of a rotating BEC with attractive self-interactions beginning at $\rho = 3\rhoeq$ with 
$\dot\rho(0)=0$. We have taken $f=n/\nc = 0.8$ and $b = 10^4$. The dashed line indicates the equlibrium radius.}
\label{ColRebnegative}
\end{minipage}
\end{figure}

We have seen that there is no local minimum with a positive $\rho$ for clouds with attractive self interactions unless the number 
of bosons is below some critical value, $\nc$, which is determined by the strength of the interactions. One can understand this as 
saying that beyond $\nc$ the attractive inter-particle energy is sufficient to overcome the quantum pressure and 
the condensate implodes.  When a local minimum exists, the equilibrium radius, $\rhoeq$ depends on the actual number of bosons, $n$, 
present (see figure \ref{EffPotnegative}). The equilibrium energy, $E_\text{eq} = V(\rhoeq)$ and also depends on the number of bosons. 
For an example, we have taken $b\approx 10^4$ and $n/\nc = 0.8$ ($\rhoeq=0.583$, $b\rhoeq \sim 6\times 10^3\gg 1$) and integrated 
\eqref{eomrho} to obtain a snapshot of the collapse process beginning at $\rho(0)=3\rhoeq$ with $\dot\rho(0) = 0$ (see figure 
\ref{ColRebnegative}). As is to be expected because of the negative pressure and weakend gravitational field, the collapse of the 
halo is extremely slow, taking on the order of 10 by to cross equilibrium.

The first integral of the motion \eqref{eom2} allows us to determine the frequency of small oscillations about equilibrium. 
They occur with characteristic period
\beq
T = 2\pi \sqrt{\frac 52}\frac{\hbar b^2}{mc^2} \sqrt{\frac{m_\text{eff}(\rhoeq)}{\oV_\text{eff}''(\rhoeq)}}.
\eeq
This gives approximately 1.3 by for repulsive self interactions with the above parameters and roughly four times longer
for attractive self interactions. This discrepancy is to be expected as the contribution to the BEC pressure from repulsive 
interactions strengthens the gravitational attraction and weakens it when the interactions are negative.

\section{Summary}

While the standard CDM paradigm for DM does well on very large scales, few of its predictions on scales less 
that $\sim$ 50 kpc have been successful. Chief among these are the absence of DM cusps at the centers of galaxies 
and the absence of an abundance of low mass and massive sub-halos predicted by the model. It is therefore worthwhile 
to analyze alternative models that behave like CDM on large scales but are more in keeping with observations on 
smaller scales. One such model proposes that at least a component of DM may consist of ultralight condensed bosons.

We have analyzed a model for gravitationally bound BEC dark matter vortices based on the Gross-Pitaevskii equation 
with quartic interactions, taking into account the effects of rotation. To include rotation, we introduced an axisymmetric 
background differing only weakly from flat space. By analyzing the weak field Einstein equations 
in the harmonic gauge, we determined the form of the metric and set up the non-relativistic action for the combined 
matter and gravitational fields. From this action we determined the equations governing the system and analyzed the 
criteria for stability. 

In order to get a better feeling for stable configurations of the BEC, we used the variational approach. With a standard 
Gaussian Vortex ansatz for the trial wave function, in which the BEC halo is assumed to have a radial dependence of $r/R$,  we 
obtained estimates for small and large condensates. The number of bosons in the condensate and its size depend 
only on the mass of the bosons and the strength of the self-interactions. As in the case of non-rotating BECs, two distinct 
cases arise with the votex ansatz as well. When the self interactions are attractive, a local minimum of the effective 
potential energy is present only when the number of bosons in the cloud is below a certain critical value. There is also a local 
maximum of the effective potential energy. If the number of particles exceeds the upper limit or when the total energy 
is greater than the maximum of the potential energy, it appears that the cloud would collapse into a black hole. However,
one cannot be sure of this outcome due to the non-relativistic, linear approximation used in this paper. Harko \cite{h14}, 
Levkov et. al. \cite{lpt17} and Eby et. al. \cite{emw17} have proposed, in an alternative scenario, that as the central 
density grows and exceeds a certain critical value a fraction of 
the bosons will get expelled from the condensate, which will then stabilize. When the interaction strength is large enough, 
we obtained a simple relation between the critical (maximum) mass and the critical (minimum) radius of the cloud on the one 
hand and the mass and the coupling strength of the bosons on the other. Even in the non relativistic limit examined here, 
no such upper limit on the number of bosons (except the Schwarzschild limit) is apparent when the self interactions are 
repulsive. In this case, there is always a global minimum of the effective potential energy. 

For example we considered attractive interactions, inputting the values of the interaction strength, $b \sim 10^7$ and 
mass, $m \sim 10^{-5}$ ev, for the QCD axion. We obtained stable condensates approximately 400 km in radius having a mass 
of approximately $10^{19}$ kg. On the other hand, for sufficiently light bosons it is possible to achieve BECs of galactic 
size. We found that taking the boson mass $m \sim 10^{-24}$ ev along with an interaction strength of $b 
\sim 10^4$ yielded a cloud with an outer radius (the radius within which 99\% of the matter is contained) of roughly 
50 kpc. Because of the density profile of a vortex, the gravitational field due to the BEC inside the core is outward 
directed up to a distance of about 17\% of the outer radius. In this example, the central region is roughly 6-8 kpc in radius 
and dominated by ordinary (baryonic) matter, with the BEC taking over the gross gravitational dynamics beyond this 
distance.

We also analyzed the dynamics of rotating BECs by considering time dependent wave functions. The time dependence was introduced 
by employing a collective coordinate description in terms of the condensate radius, $R=R(t)$. Equations for the evolution of 
$R(t)$ were obtained from an effective action, achieved by integrating the action for the combined BEC and gravitational field. 
The choice of the trial wave function is not unique of course and the extent to which the results obtained with 
different trial wave functions differ qualitatively from one another is a topic for future investigation. In our ansatz there 
is just one free parameter and the system is one dimensional. The Poisson equations can be solved 
exactly and the gravitational potentials of the halo can be evaluated in analytical form. The motion of the condensate then 
becomes analogous to the motion of a single particle of variable mass in an effective potential. In an ansatz with multiple 
parameters one may expect multiple coupled equations, which could be considerably more difficult to solve. We showed from the 
dynamics that the equilibrium conditions are identical to the ones obtained earlier in the static case. Moreover, in both the 
attractive and repulsive cases, collapse from a diffuse state into the equilibrium state is a process that takes billions of 
years. We have also examined the time scale for small oscillations about equilibrium and found it to be, not surprisingly, 
on the same order of magnitude.

There are several aspects of this description that have not been addressed in this work. For one, our model does not include a 
microscopic mechanism for varying the number of particles in the condensate, so we cannot say what happens, for example, 
if a BEC near its critical mass is immersed in a cloud of free bosons. Again, our model does not include damping, so a BEC 
will oscillate about its equilibrium radius essentially forever. We will consider damped BEC models in a future work. It 
would also be interesting to examine what happens when several types of BECs or even a single boson with multiple accessible 
states share a single gravitational well. 
\bigskip\bigskip

\centerline{\bf Acknowledgments}
\bigskip

\noindent L. C. R. W. thanks J. Eby, M. Leembruggen, M. Ma  and P. Suranyi for discussions.

\end{document}